\documentclass[apj]{emulateapj}

\newcommand{\deltE}{\Delta\kern-1ptE}

\input epsf

\usepackage{longtable}
\usepackage{amssymb}
\usepackage{graphicx}
\usepackage{epsfig}
\usepackage{epsf}

\slugcomment{To be submitted to ApJ}
\shorttitle{X-ray irradiated protoplanetary disk atmospheres {\sc ii}}
\shortauthors{Ercolano et al.}
\begin{document}
\title{X-ray irradiated protoplanetary disk atmospheres II:\\ 
Predictions from models in hydrostatic equilibrium}
\author{Barbara Ercolano$^1$, Cathie J. Clarke$^1$, Jeremy J. Drake$^2$}
\affil{$^1$Institute of Astronomy, University of Cambridge, \\
Madingley Rd, \\Cambridge, CB3 OHA, UK\\
$^2$Harvard-Smithsonian Center for Astrophysics, MS-67
\\ 60 Garden Street, \\ Cambridge, MA 02138, USA}

\begin{abstract}
We present new models for the X-ray photoevaporation of circumstellar discs
which suggest that the resulting mass loss (occurring mainly
over the radial range $10-40$ AU) may be the dominant dispersal
mechanism for gas around low mass pre-main sequence stars, contrary
to the conclusions of previous workers. Our models
combine use of the MOCASSIN Monte Carlo radiative transfer code
and a self-consistent solution of the hydrostatic structure of the 
irradiated disc. We estimate the resulting photoevaporation rates assuming
sonic outflow at the surface where the gas temperature equals the local
escape temperature and derive  mass loss rates of $\sim 10^{-9} M_\odot$ 
yr$^{-1}$, 
typically a factor $2-10$ times lower than the corresponding rates in our previous work
(Ercolano et al., 2008) where we did not adjust the density structure of
the irradiated disc. The somewhat lower rates, and the fact that mass loss
is concentrated towards slightly smaller radii, result from the puffing up
of the heated disc at a few AU which partially screens the disc at tens of
AU. Our mass loss fluxes  agree with those of Alexander et al. (2004)
but  we differ with Alexander et al. in our assessment of the overall
significance of X-ray photoevaporation, given the large disc radii
(and hence emitting area) associated with X-ray driven winds.
Gorti and Hollenbach (2009), on the other hand, predict considerably lower mass
loss fluxes than either Alexander et al., (2004) or ourselves and we discuss
possible reasons for this difference. We highlight the fact that X-ray 
photoevaporation has two generic advantages for disc dispersal
compared with photoevaporation by extreme ultraviolet (EUV) photons that
are only modestly beyond the Lyman limit: the demonstrably large X-ray
fluxes of young stars even after they have lost their discs and the
fact that X-rays are effective at penetrating much larger columns of material close
to the star.   We however stress that our X-ray driven mass loss rates are 
considerably more  uncertain than the corresponding rates for
EUV photoevaporation (around $10^{-10} M_\odot$ yr$^{-1}$) and that this
situation will need to be remedied through future radiation
hydrodynamical simulations. 

\end{abstract}

\keywords{}

\section{Introduction}\label{s:intro}

The study of the structure and evolution of protoplanetary disks is currently 
an important area of astronomy as it can provide insights into both star
and planet formation. In recent years there has been considerable debate
about the final destination of disc material - do discs disperse through a
mixture of accretion onto the star and internal consumption by conversion
into planets, or are they instead dispersed by an extrinsic agent which
can operate in competition with planet formation? Obviously, the answer
to this question relates to the lifetime of circumstellar discs
and the efficiency of planet formation around stars of various masses.

  Since disc dispersal by central wind stripping is not very
    efficient (Matsuyama, Johnstone \& Hollenbach, 2009) and since encounters
    with passing stars is ineffective 
in all but the densest environments (Scally \& Clarke 2001, 
Pfalzner et al., 2006), and since likewise the effect of
photoevaporation by neighbouring OB stars requires a very rich cluster
environment (Johnstone et al., 1998), the most likely extrinsic agent of
disc dispersal is via photoevaporation by energetic (ultraviolet or X-ray)
radiation of the central star.  Indeed the detection of Ne~{\sc ii} emission in the 
spectra of e.g Pascucci et al. (2007) and Herczeg et al. (2007) is widely
interpreted as evidence for
disc irradiation by high energy photons and potentially
accompanying  photoevaporation. Opinion is however currently divided
as to whether this is best explained in terms of irradiation by X-rays (e.g. 
Glassgold 2007, Ercolano 2008b,  Meijerink 2008) or ionising ultraviolet
(EUV) radiation (Gorti \& Hollenbach 2008, Alexander 2008). 

 The X-ray properties of low mass pre-main sequence stars are well
 determined (in contrast to the case for the EUV radiation from these
 stars), with flux levels that are, if anything, somewhat higher in
 the case of Weak Line (discless) T-Tauri stars than in their disc
 bearing (Classical T Tauri star) counterparts (Preibisch et al.,
 2005, Feigelson et al., 2007): although this may result from
 absorption of some of the X-ray emission in accretion flows close to the star
(Gregory et al., 2007, but see also Drake et al., 2009), it at least
 demonstrates that the X-rays remain strong throughout and after the
 disc bearing stage and are therefore, in principle, available for
 disc dispersal. The corresponding observational situation in the EUV
 case is much less clear, given the lack of suitably 
high quality ultraviolet spectra, particularly in Weak Line T Tauri stars (Alexander, Clarke \& Pringle, 2005; Kamp \& Sammar, 2004). 
 
  Another potential drawback to EUV photoevaporation is that, even if one
has an accurate estimate of the EUV output of the star, it does not
necessarily follow that these photons can reach the disc, since photons
near the Lyman limit are particularly susceptible to absorption by even 
small columns of neutral material close to the star. This represents
a possible counter-argument to Ne~{\sc ii} emission being excited by EUV
radiation in most cases. It is usually therefore argued (e.g. Gorti \&
Hollenbach, 2009) that EUV photoevaporation cannot be effective until 
late times when the subsiding of both accretion and outflow activity
should allow the disc to be exposed to ionising photons. From this
point onwards, the theory is well developed (Hollenbach et al., 1994,
Clarke et al., 2001, Alexander et al., 2006a,b), with a predicted 
photoevaporation rate of $10^{-10} M_\odot$ yr$^{-1}$, for an EUV
  luminosity of $\sim$10$^{41}$s$^{-1}$ (the rate goes as the square root of the assumed luminosity), and the interplay 
between photoevaporation and accretion producing a characteristic
evolutionary sequence involving the creation and rapid outward expansion
of an inner disc hole.  Thus EUV photoevaporation provides an
alternative to planet formation as the origin of inner hole (transition
disc) systems (see e.g. Najita et al., 2007).

  Notwithstanding this success of EUV photoevaporation at late times
in removing the last few Jupiter masses of gas from the disc, there
has also been some discussion  of whether other radiation sources, such as X-rays
or non-ionising (FUV) ultraviolet photons, might not be much more
effective disc photoevaporation agents at early times 
and might indeed be capable of removing much larger quantities of
disc gas. Gorti \& Hollenbach (2009) investigated this issue and argued
that FUV photoevaporation is indeed a very significant effect at early
times when accretion produces a strong 
FUV flux: 
their predicted FUV mass loss rates of $> 10^{-8} M_\odot$ yr$^{-1}$ exceed
EUV rates by  more
than two orders of magnitude. They however concluded, by contrast, that
X-rays, in the absence of FUV radiation, are considerably less important
even than the EUV, a conclusion also shared by Alexander et al. (2004).

 Another recent study of X-ray photoevaporation was provided by  Ercolano  et al
2008b, henceforth Paper~I. While Paper~I mainly aimed at understanding the
thermochemical structure of the irradiated regions of the disk and at
identifying gas-phase emission line diagnostics (see also Ercolano, Drake \&
Clarke 2008c), it also yielded an estimate for the X-ray photoevaporation
rate that was considerably higher than those found by Gorti \&
Hollenbach (2009) and Alexander et al (2004), being around $10^{-8} M_\odot$ yr$^{-1}$. 
This estimate was  however  very preliminary, due
not only to the general difficulty of estimating mass loss rates from
static models (see Section 6),   but on account of the fact that 
the disc's density structure was not modified in response to heating of
its upper layers. Instead it was  assumed that the disc adopted the fixed
structure that it would have in the case that its temperature
structure was controlled by full 
thermal coupling between dust and gas (d'Alessio et al., 1998). This
is clearly inconsistent given that the gas and dust temperatures in the
X-ray irradiated disc turn out to be completely decoupled down to a
column depth of $\sim$5$\times$10$^{21}$\,cm$^{-2}$.
In contrast,  Gorti \& Hollenbach (2009) and
Alexander et al. (2004) both iterated to a self-consistent
hydrostatic density structure for the irradiated disc. Since these authors
had previously concluded that X-ray photoevaporation is a minor effect, it would
appear necessary to check how a similar iteration onto a hydrostatic
structure would affect the initial estimate of Paper~I. The reason
for pursuing a study that is broadly similar to those of Gorti \&
Hollenbach (2009) and Alexander et al. (2004), is that there are a number of differences
in the way that the radiative transfer is treated in the three works.
We employ a fully three dimensional Monte Carlo approach
to the radiative transfer through dust and gas, and, while  - in the
present paper - we  iterate on the
density profile at each cylindrical radius, the final radiation field
and temperature structure represents the thermal equilibrium condition for
a given density field, without any simplifying assumptions about the
geometry of the resulting radiative flux vectors. The other two works
instead employ methods that are pseudo-one dimensional in that they integrate
along a range of lines of sight from the X-ray source. There are also
some other differences concerning the assumed spectrum of the X-ray source
and some details about the detailed interaction between X-rays and gas
that are discussed in more detail in Section 5.

The calculations presented here thus aim specifically at
providing an improved estimate of  X-ray photoevaporation rates, together
with a comparison with the results of using irradiating spectra that
also include an  EUV component. 
For computational tractability, only the inner 50AU of the disk, from which significant photoevaporation is found, are
considered here and therefore a full emission line spectrum is not
available. This will be investigated in future work. 

A description of the computational methods is given in Section~2.
Our improved photoevaporation rates are presented and
discussed in Section~3, while Section~4 describes the resulting
thermal and ionisation 
structures. Section ~5 analyses differences in methods and results compared
with previous studies and Section ~6 sets out the main uncertainties
involved in estimating mass loss rates from hydrostatic models.
Section~7  summarises our main results.

\section{Photoionisation models in hydrostatic equilibrium}
\label{s:model}

We have used a modified version of the 3D photoionisation and
dust radiative transfer code {\sc mocassin} (Ercolano et al., 2003; 2005;
2008a).  This code uses a Monte Carlo approach to the three-dimensional
transfer of radiation, allowing for the self-consistent transfer of
both the stellar (primary) and diffuse (secondary) components of the
radiation field. Details of atomic data and physical processes used
for the photoionisation calculations are given by Ercolano et
al. (2003, 2008a) and Ercolano \& Storey (2006). The main changes to
the thermal balance and dimensional setup of the code have been  
described in detail in Paper~I and are briefly summarised here. 
(i) A treatment of viscous heating is included using the formulation for a
thin disk by Pringle (1981). 
(ii) Gas cooling by collisions of grains with
a mixture of atomic and molecular hydrogen is calculated using the
method of Hollenbach \& McKee (1979), also discussed by Glassgold et
al. (2004). 
(iii) {\sc mocassin} was adapted to allow for two dimensional
  simulations in order to exploit the symmetry of the system. 

In Paper~I, viscous accretion heating was included using the
standard treatment for a thin disk (Pringle 1981) with viscosity
parameter $\alpha$~=0.01.  This corresponds to an accretion rate of
the order of 10$^{-8}M_{\odot}/yr$. This is rather high, in particular
as we are interested in the effects of X-ray photoevaporation in later
stages of disk evolution when accretion has turned off. This is
achieved in our models by setting $\alpha$~=10$^{-10}$ which
effectively renders accretion heating negligible.
We have also improved the model in Paper~I by including excitation of
C~{\sc i} and O~{\sc i} fine structure lines by collisions with
neutral hydrogen (only collisions with electrons were previously
included, see Ercolano, Drake \& Clarke, 2008 for more details).

\subsection{Hydrostatic Equilibrium}
\label{s:hydro}

We have further modified the {\sc mocassin} code to include a
self-consistent calculation of the 1D vertical 
hydrostatic equilibrium structure (HSE) of the X-ray heated disk.  We used
the methods described by Alexander, Clarke and Pringle (2004, ACP04) 
to update the disk density structure after convergence of the thermal and
ionisation structure based on the previous HSE iteration. 

Our initial density distribution is the same as that used in
Paper~I and it consists of the disk model structure calculated by
D'Alessio et al. (1998), selected to be that which best fits the
median SED of T~Tauri stars in Taurus (d'Alessio, 2003). This model
was calculated under the assumption of full thermal 
coupling between dust and gas and featured an irradiating star of
mass 0.7~M$_{\odot}$, radius 2.5~R$_{\odot}$, and an effective temperature of
4000~K. Disk parameters include a mass accretion rate of 10$^{-8~}$ M$_{\odot}$ yr$^{-1}$ and a viscosity parameter $\alpha$~=~0.01. The
total mass of the disk is  0.027~M$_{\odot}$ and the outer radius of
$\sim$500~AU. The surface density in the disk is inversely proportional to the
cylindrical radius. 

We perform HSE calculations in the vertical
direction, using as a lower boundary the point on the z-axis where the
gas and dust temperatures become decoupled. As in Paper~I, we
keep the dust temperatures fixed to the values calculated by 
D'Alessio et al. (2001) given that without treating the stellar or
interstellar optical and far-UV fields, the dust temperatures we would
obtain from our model would be incorrect. Calculating the hydrostatic
equilibrium from the decoupling point has the advantage of keeping a
realistic density structure below this point even when only X-ray
irradiation is considered (i.e. not treating the photospheric emission
of the pre-main sequence star). 

The density structure from the lower boundary point is calculated
using a slightly modified version of equation~A9 of ACP04, as shown below.
At a given cylindrical radius, the number density of the gas as a
function of height above the midplane, $z$, is given by

\begin{equation}
\label{eq:hydro2}
n(z) = n_R \frac{T_R}{T(z)} exp\left[-\frac{m_H}{k} G M_* \beta(z)\right]
\end{equation}

\noindent where $n_R$ and $T_R$ are density and temperature at the lower
boundary, $M_*$ is the mass of the star, $G$ is the gravitational
constant, $m_H$ is the mass of the
hydrogen atom and $k$ is the Boltzmann constant. $\beta(z)$ is calculated
as follows

\begin{equation}
\label{eq:hydro}
\beta(z) = \int_{z_R}^z \frac{\mu(z') z'}{T(z') (z'^2 + R^2)^{\frac{3}{2}}}dz'
\end{equation}

\noindent where $\mu(z)$ is the mean molecular weight of the gas at $z$ and $R$ is the cylindrical radius. 

Equations~\ref{eq:hydro2} and \ref{eq:hydro} differ from equation~A9 of
ACP04 in that we do not assume z $<<$ R. We note, however, that
a full treatment 
would consist of a two-dimensional HSE in
the radial and vertical directions -- with gravity balanced by
pressure in the vertical direction and with gravity balanced by a
combination of pressure and centrifugal forces in the radial
directions. In the case of an irradiated disk with significant vertical 
extent the pressure
terms are very complicated and a full solution is not attempted in
this paper. It is indeed unclear whether a fully two-dimensional
hydrostatic solution would improve our calculations, particularly in
the upper layers of the disk, where the assumption of hydrostatic
equilibrium is altogether poor given that material is likely to be
flowing in a photoevaporative wind. However a full hydrodynamical flow
solution is beyond the scope of this paper. 

\subsection{The dust model}

 Although we do not calculate the dust temperatures, the effects of
 gas-dust interactions on the temperature structure of the gas, are
 still taken into account in our gas thermal balance as well 
 as in the radiative transfer, where the competition between dust and
 gas for the absorption of radiation is properly treated (see Ercolano
 et al., 2005). Our
 dust treatment consists of a standard MRN-type model (Mathis, Rumpl \&
Nordsiek, 1977) with minimum and maximum grain radii $a_{min}= 0.005\mu$m and
$a_{max}=0.25\mu$. The disk structure of d'Alessio et al. (1998), which
provides our starting density distribution, employs a bimodal dust
distribution, where atmospheric dust follows the standard MRN model and
interior dust consists of larger grains with a size distribution
still described by a power law of index of $-3.5$, 
but with $a_{min}= 0.005\mu$m and
$a_{max}=1mm$. The transition between atmospheric and interior dust
occurs at an height of 0.1 times the midplane gas scale height. As we
are mainly interested in regions well above the transition point,
we have chosen for simplicity to use atmospheric dust everywhere.
Following d'Alessio et al. (2001) the dust to gas mass ratio of
graphite is 0.00025 and that of silicates is 0.0004.

High energy dust absorption and scattering coefficients are calculated
from the dielectric constants for graphite and silicates
of Laor \& Draine (1993), which extend to the X-ray domain.  Spherical
grains are assumed and we use the
standard Mie scattering series expansion for
complex refractive indices, $x|m| < 1000$, where
$x~=~2~\cdot~a/\lambda$ is the scattering parameter (see Laor \& 
Draine, 1993). For $x|m| > 1000$ and $x|m-1| < 0.001$ we use
Rayleigh-Gans theory (Bohren \& Huffman, 1983), and for $x|m| > 1000$
and $x|m-1| > 0.001$ we use the treatment specified by Laor \& Draine
(1993), which is based on geometric optics.  

\subsection{T-Tauri EUV-X-ray Spectra}

We illuminate the disk using synthetic X-ray spectra representative of
the high-energy emission from a typical T Tauri star corona.  This
coronal X-ray emission arises from a collision-dominated
optically-thin plasma.  

\begin{figure}
\plotone{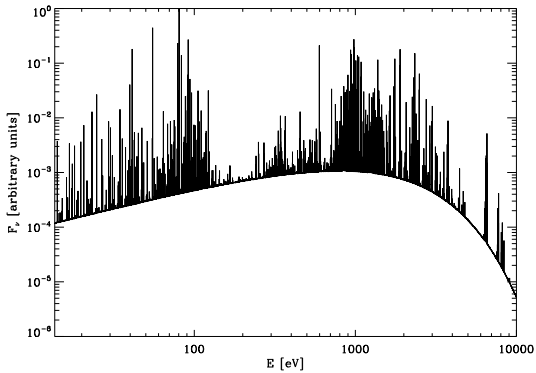}
\caption[]{Isothermal model spectrum for Log(T$_X$) = 7.2. }
\label{f:sp1}
\end{figure}

We first assumed a single X-ray temperature, $\log(T_x)=7.2$, and a
number of luminosities spanning 2 orders of magnitudes around our base
luminosity of $L_X$(0.1-10~keV)$=2\times10^{30}$~erg/sec (Model
XS0H10Lx1). The
labelling scheme of our models is explained in Section~2.4. The bulk
of the luminosity in these models is emitted in 
the 1-2~keV region as shown in  Figure~\ref{f:sp1}. The same
unabsorbed isothermal 
model spectrum for Log(T$_X$) = 7.2 was also used in Paper~I and 
we refer to that work for a more detailed description of the
computational methods and atomic data used.  

While such a model provides an accurate description of the X-ray
spectrum, it represents only a lower limit to the unabsorbed flux at
EUV wavelengths ($\sim 100-912$~\AA) where cooler plasma can
contribute significant flux.  In Paper~I we argued that for most of a
disk's lifetime EUV photons would not easily reach the disk due to
absorption from circumstellar material and restricted our
investigation to the X-ray dominated irradiation spectrum above. Here
we expand our study to explore the effects that different amounts of
EUV irradiation would have on a disk that is also simultaneously
irradiated by X-ray. In a companion paper (Glassgold et al., 2009 in
prep.) we study the relative importance of EUV and X-rays on
ionisation and heating rates, while here we focus on their effects on
photoevaporation.  To this aim, we employ a more sophisticated
multi-temperature plasma model developed to provide an accurate representation of
the ionizing coronal spectrum throughout the EUV-X-ray range.  We have
not included any contributions to the ionizing flux from accretion
here; this will be considered in future work.

Observed line-of-sight absorbing columns toward T~Tauri stars are
typically greater than $10^{20}$ atoms cm$^{-2}$ (e.g. Feigelson \&
Montmerle 1999), and consequently their EUV spectra are completely
obscured from our view by photoelectric absorption.  While this would
appear to be a rather serious impediment to producing a realistic
T~Tauri EUV spectrum, there are much more nearby and relatively
unabsorbed sources that provide a template: the RS~CVn class of active
binaries that comprise at least one evolved star lying on the
sub-giant or giant branch.  The coronal emission of RS~CVn binaries is
largely dominated by the evolved component that is likely to be
magnetically more similar to a T~Tauri star than are active
main-sequence stars with shallower convection zones.  Rotation periods
of RS~CVn binaries are also similar to those of T~Tauri stars and
typically lie in the range 1-10 days (e.g., Strassmeier et al. 1993,
Getman et al., 2005).

The expected resemblance of the coronal properties of RS~CVn-type
binaries and T~Tauri stars is borne out in X-ray studies.  For
example, the {\it Einstein} survey of late-type stars by Schmitt et
al. (1990) obtained plasma temperatures for RS~CVns in the range $7
\la \log T \la 7.8$ with a mean of order $\log T \sim 7.3$, and 
X-ray luminosities in the range $10^{30} \la L_X \la
10^{31.8}$~erg~s$^{-1}$.  These ranges are very similar to those found
for T~Tauri stars of the Orion Nebular Cluster (ONC) in the mass range
1-2$M_{\odot}$ by Preibisch et al. (2005).  On this basis, the 
use of RS~CVn EUV spectra as a proxy for the coronal spectra of 
T~Tauri stars seems well-justified.

\begin{figure}
\plotone{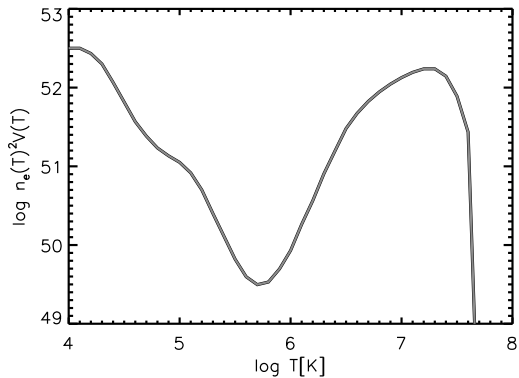}
\caption[]{Emission measure distributions as a function of temperature.}
\label{f:emd}
\end{figure}

Detailed studies of the outer atmospheric temperature structure of
RS~CVn-type binaries also finds similarity in the transition regions
and lower coronae of different stars.  Sanz-Forcada et al. (2002)
studied four active RS~CVns in detail from UV to the shortest EUV
wavelengths using spectra obtained from the {\it International
Ultraviolet Explorer} (IUE), the {\it Extreme Ultraviolet Explorer}
(EUVE) and the {\it Orbiting and Retrievable Far and Extreme
Ultraviolet Spectrometer} (ORFEUS).  They derived emission measure
distributions as a function of temperature for the range $\log
T=4.5$--7.2 and we use a representation of these for the emission
measure distribution adopted as the basis of our T~Tauri model.  This
emission measure distribution is illustrated in Figure~\ref{f:emd}.
The Sanz-Forcada et al. (2002) emission measures are rather
ill-constrained for temperatures $\log T > 7$ owing to the paucity of
spectral line diagnostics in the EUV range at these temperatures.  For
this range we have adopted a smooth function that peaks at $\log
T=7.25$ and drops off fairly sharply for temperatures $\log T > 7.5$.
The temperature of the peak is based on the two-temperature
optically-thin thermal plasma model fits to the large sample of {\it
Chandra} ONC T~Tauri spectra by Maggio et al. (2007).

\begin{figure}
\plotone{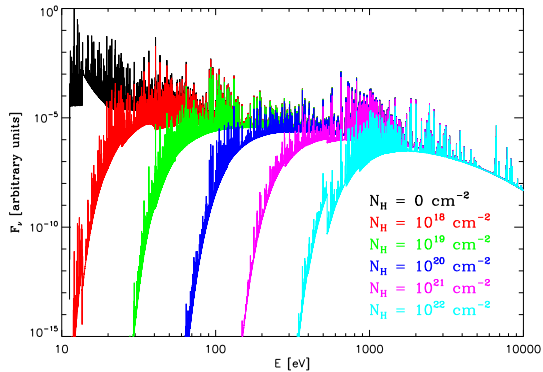}
\caption[]{Model coronal spectra based on the emission measure
  distribution shown in Figure~\ref{f:emd}. The black line represents
  the unattenuated spectrum, while coloured lines refer to spectra
  transmitted through semi-neutral columns of the values shown. For
  clarity, the spectra are shown in arbitrary units and are not
  normalised to the X-ray luminosity values used in the modelling. }
\label{f:jspectra}
\end{figure}

Coronal spectra based on the emission measure distribution in
Figure~\ref{f:emd} were computed as described in Paper~I using the
{\sc PINTofALE} IDL software suite\footnote{PINTofALE 
is freely available from http://hea-www.harvard.edu/PINTofALE}
(Kashyap \& Drake 2000), using line and continuum emissivities from the
CHIANTI compilation of atomic data (Landi \& Phillips 2006, and
references therein), ion populations from Mazzotta et al. (1998), and
the solar chemical composition of Grevesse \& Sauval (1998).
Figure~\ref{f:jspectra} shows the
resulting EUV+X-ray spectrum produced (black line);  the emergent
luminosity is roughly the same in the EUV (13.6$\leq \nu \leq$100 eV) and 
X-ray regions ($\nu > 100 eV$).

In order to account for absorption of coronal EUV and X-ray photons in
the immediate local stellar and circumstellar environment, disk
irradiation calculations were also performed for spectra attenuated by
photoelectric absorption corresponding to neutral hydrogen column
densities in the range $N_H=10^{18}$--$10^{22}$~cm$^2$.  The gas
transmittance was computed using the cross-sections compiled by Wilms,
Allen, \& McCray (2000) as implemented in {\sc PINTofALE}.
 The effects of different 'pre-screening' columns on the radiation impinging
 on the disk 
are shown by the coloured lines in Figure~\ref{f:jspectra}, where the
black, red, green, blue, magenta and cyan lines show the transmitted spectrum
through columns of 0., 10$^{18}$, 10$^{19}$, 10$^{20}$, 10$^{21}$ and
10$^{22}$~cm$^{-2}$, respectively. The corresponding models are
labeled FS0H2, FS18H2, FS19H2, FS20H2, FS21H2, FS22H2. 
For the disk photoionisation calculations, spectra were normalised to 
an X-ray luminosity $L_X(0.1-10~keV) = 2 \cdot 10^{30} erg/sec$. 

Finally, in Paper~I, we assumed the height of the stellar 
irradiating source 
to be 10 stellar radii (R$_*$) above the disk mid-plane, as was adopted 
in earlier studies 
(e.g.\ Glassgold et al., 1997, Igea \& Glassgold 1999) 
This represents quite a large scale height for coronal emission, and the
source of quiescent X-rays is likely to reside much closer to the star.  
Indeed, Flaccomio et al.\ (2005) found evidence for rotationally 
modulated  X-rays from T~Tauri stars in the Orion Nebular Cluster, 
indicating that the
dominant emission from those stars lies at scale heights of order the stellar
radius or below.  Gregory et al.\ (2006) also find that simple 
extrapolations of T~Tauri surface magnetograms 
suggest compact coronal structures.
We therefore adopt 
here a height of 2~R$_*$ for the X-ray source, corresponding to the
X-ray emission originating from a point located 1~R$_*$ above the
stellar pole.  We also investigate the effect of this scale height
choice and find that changing the source height
between 2 and 10 R$_*$ has relatively little influence on the results.

\subsection{Model specification}

Table~\ref{t:mods} contains a summary of the parameters used for the
models presented in this paper. In general models are labeled as
nSsHhLxl, where 'n' is 'F' for models using EUV+X-ray irradiating
spectra (Figure~\ref{f:jspectra}) and 'X' for models using
the single temperature Log(T$_X$)~=~7.2 spectrum. 's' indicates the
pre-screening column and it is 0 for no pre-screening and 22 for a pre-screening
column of 10$^{22} cm^{-2}$. 'h' indicates the height of the irradiating
source which is 2 or 10 (stellar radii). 'l' indicates the X-ray
luminosity used, e.g. 01, 1, 10 for L$_X ~=~2\cdot10 ^{29}$,
2$\cdot10^{30}$, 2$\cdot10^{31}$, respectively.  

\begin{table*}
\begin{center}
\caption[]{Model input parameters and predicted mass loss rates from
  photoevaporation. 'Spectrum' refers to the 
  irradiating spectrum, see text in Section~2.3. 'Column' is the
  column density of circumstellar screening material (pre-screening column). 'Height' is the
  height of the irradiating source.  $L_X$ is the X-ray luminosity
  defined from 100 eV $<$ E $<$ 10 keV. 'Hydro' refers to whether the
  models are in hydrostatic equilibrium (TRUE) or a fixed density 
structure was
  used (FALSE)}
\label{t:mods}
\begin{tabular}{lcccccc}
\hline
               & Spectrum   & Column      & Height    & $L_X$      & Hydro &  \.M    \\
               &            & $[cm^{-2}]$ &  $[R_*]$  & $[erg/sec]$ &      & [M$_{\odot}$/yr]  \\
\hline
XS0H2Lx1       & X-ray      &    0.        &   2      &  2.e30     & TRUE  & 2.0e-9  \\
XS0H10Lx1      & X-ray      &    0.        &  10      &  2.e30     & TRUE  & 1.2e-9  \\
XS0H10Lx01     & X-ray      &    0.        &  10      &  2.e29     & TRUE  & 2.4e-10 \\
XS0H10Lx02     & X-ray      &    0.        &  10      &  4.e29     & TRUE  & 4.5e-10 \\
XS0H10Lx04     & X-ray      &    0.        &  10      &  8.e29     & TRUE  & 7.0e-10 \\
XS0H10Lx08     & X-ray      &    0.        &  10      &  1.6e30     & TRUE  & 1.1e-9 \\
XS0H10Lx2      & X-ray      &    0.        &  10      &  4.e30     & TRUE  & 2.2e-9  \\
XS0H10Lx4      & X-ray      &    0.        &  10      &  8.e31    & TRUE  & 4.0e-9  \\
XS0H10Lx10     & X-ray      &    0.        &  10      &  2.e31     & TRUE  & 5.9e-9  \\
XS0H10Lx20     & X-ray      &    0.        &  10      &  4.e31     & TRUE  & 1.1e-8  \\
FS0H10Lx1      & X-ray+EUV  &    0.        &  10      & 2.e30      & TRUE  & 3.5e-9 \\
FS0H2Lx1       & X-ray+EUV  &    0.        &   2      & 2.e30      & TRUE  & 4.5e-9 \\
FS18H2Lx1      & X-ray+EUV  & 10$^{18}$    &    2     & 2.e30      & TRUE  & 4.5e-9 \\
FS19H2Lx1      & X-ray+EUV  & 10$^{19}$    &    2    &   2.e30     & TRUE  & 4.2e-9 \\
FS20H2Lx1      & X-ray+EUV  & 10$^{20}$    &    2     & 2.e30      & TRUE  & 4.0e-9 \\
FS21H2Lx1      & X-ray+EUV  & 10$^{21}$    &    2     & 2.e30      & TRUE  & 2.7e-11 \\
FS22H2Lx1      & X-ray+EUV  & 10$^{22}$    &    2     &  2.e30     & TRUE  & --      \\
XS0H2Lx1\_FG   & X-ray      &    0.        &   2      &  2.e30     &FALSE  & 1.1e-8 \\
XS0H10Lx1\_FG  & X-ray      &    0.        &  10      & 2.e30      &FALSE  & 8.1e-9 \\
FS0H10Lx1\_FG  & X-ray+EUV  &    0.        &  10      & 2.e30      &FALSE  & 2.2e-8 \\
FS0H2Lx\_FG    & X-ray+EUV  &    0.        &   2      & 2.e30      &FALSE  & 2.2e-8 \\
FS18H2Lx1\_FG  & X-ray+EUV  & 10$^{18}$    &    2     & 2.e30      &FALSE  & 1.5e-8 \\
FS19H2Lx1\_FG  & X-ray+EUV  & 10$^{19}$    &    2    &   2.e30     &FALSE  & 1.3e-8 \\
FS20H2Lx1\_FG  & X-ray+EUV  & 10$^{20}$    &    2     & 2.e30      &FALSE  & 7.7e-9 \\
FS21H2Lx1\_FG  & X-ray+EUV  & 10$^{21}$    &    2     & 2.e30      &FALSE  & 1.3e-9 \\
FS22H2Lx1\_FG  & X-ray+EUV  & 10$^{22}$    &    2     &  2.e30     &FALSE  & -- \\

\hline
\end{tabular}
\end{center}
\end{table*}

We have adopted the following elemental abundances, given
as number densities with respect to hydrogen, He/H~=~0.1,
C/H~=~1.4$\times$10$^{-4}$, N/H~=~8.32$\times$10$^{-5}$,
O/H~=~3.2$\times$10$^{-4}$, Ne/H~=~1.2$\times$10$^{-4}$,
Mg/H~=~1.1$\times$10$^{-6}$, Si/H~=~1.7$\times$10$^{-6}$,
S/H~=~2.8$\times$10$^{-5}$. These are solar abundances (Asplund et
al. 2005) depleted according to Savage \& Sembach (1996). 

Given the differences in the code and input parameters from those used
in Paper~I, we also produced a set of fixed density structure 
models in order to  
compare with the HSE and isolate the
effects of the latter. This is made necessary particularly by the fact
that the HSE models are likely to overestimate the
densities in the upper layers (which would be in reality part of a
photoevaporative flow), hence fixed density structure models, which underestimate
the density of the upper layers, can be used to provide an educated
guess of the uncertainties in the results.  ``Fixed grid'' models are
listed in Table~\ref{t:mods} with the suffix 'FG'.

\section{X-ray photoevaporation rates and timescales for disk dispersal}
\label{s:disp}

We estimate photoevaporation rates assuming that the mass loss rate
per unit surface is roughly $\dot\Sigma = \rho c_s$, where
$\rho$ and $c_s$ are, respectively, the gas density and the sound
speed evaluated at the base of the flow. The base of the flow is
defined as the location in the disk atmosphere where the gas
temperature exceeds the local escape temperature, $T_{es} = G m_H M_*
/ k R$, where $M_*$ is the stellar mass and $R$ is the radial distance. 
The total mass loss rates \.M are obtained by integrating
$\dot\Sigma$ over the disk surface. 
This simple approach, as discussed in Paper~I, introduces
significant uncertainties in the calculation of mass loss rates. These
are further discussed in Section~\ref{s:comp} and~\ref{s:lims}. 

The last column of Table~\ref{t:mods} lists the mass loss rates obtained by our models.
The sensitivity of these rates to the model parameters are discussed
in more detail below.  {\it Our general results show that soft X-rays
(roughly a few 100 eV) are most efficient at driving a
photoevaporative wind, producing a mass loss rate of the order of
10$^{-9}$ M$_{\odot}$/yr. }

\begin{figure}
\plotone{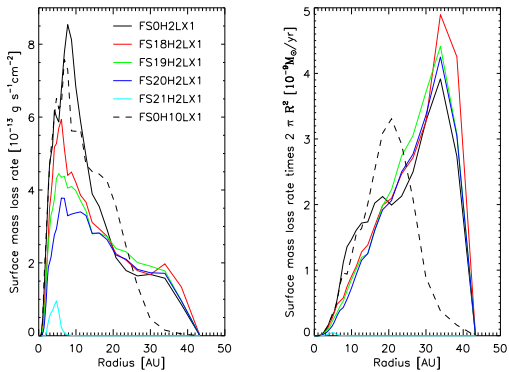}
\caption[]{Surface mass loss rates (left) and total mass loss rates
  (right) of hydrostatic equilibrium X-ray+EUV models with different
  pre-screening columns as indicated by the model labels. The solid lines
  correspond to models irradiated by a source located at a height of
  2~R$_*$, while the dashed line corresponds to an unscreened
  X-ray+EUV model irradiated by a source at 10~R$_*$.}
\label{f:sigmah}
\end{figure}

\paragraph{Illuminating spectrum and pre-screening columns}
The values in Table~\ref{t:mods} show that both for HSE and
fixed density structure
models the total mass loss rates are not very sensitive to
the inclusion of EUV ($\la$100eV) radiation.  The models show very
similar values of \.M for pre-screening columns ranging from 0. to
10$^{20} cm^{-2}$, implying that the radiation screened by columns of
up to 10$^{20} cm^{-2}$ (energies less than 100eV) is less efficient
than higher energy radiation at dispersing the disk by
photoevaporation. This can be understood by examining the surface mass
loss rates, $\dot\Sigma$, and total mass loss rates, \.M, of models
with different pre-screening columns, as shown for the hydrostatic
equilibrium models in Figure~\ref{f:sigmah}. Although $\dot\Sigma$ at
5-10AU decreases as the pre-screening column increases (see left panel of
Figure~\ref{f:sigmah}), however $\dot\Sigma$ at larger radii (20-30
AU) is roughly invariant for columns up to 10$^{20} cm^{-2}$.  Total
mass loss rates are dominated by the surface emission at larger radii,
due to the radius squared dependence of the mass loss rate integral,
and therefore they also appear to be invariant for absorbing columns
up to 10$^{20} cm^{-2}$. This is shown in the right panel of
Figure~\ref{f:sigmah} where the quantity $2 \pi R^2 \dot\Sigma$
is plotted as a function of radius, R. We plot $2 \pi R^2
\dot\Sigma$ as this quantity is generally used as a rough estimate of
\.M; in Table~\ref{t:mods}, however, we give the values of \.M
obtained by formally integrating $\dot\Sigma$ over the whole disk.

\begin{figure}
\plotone{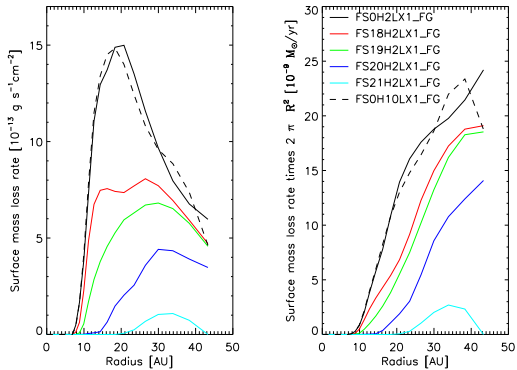}
\caption[]{Surface mass loss rates (left) and total mass loss rates
  (right) of fixed density X-ray+EUV models with different
  pre-screening columns as indicated by the model labels. The solid lines
  correspond to models irradiated by a source located at a height of
  2~R$_*$, while the dashed line corresponds to an unscreened
  X-ray+EUV model irradiated by a source at 10~R$_*$.}
\label{f:sigmanh}
\end{figure}

\begin{figure}
\plotone{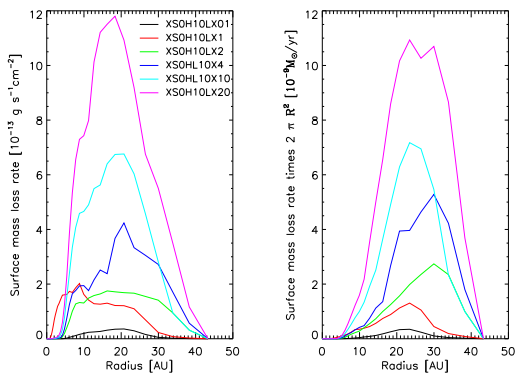}
\caption[]{Surface mass loss rates (left) and total mass loss rates
  (right) of  hydrostatic equilibrium single-temperature X-ray models
  of different luminosities as indicated by the model labels. All
  models were irradiated by a source located at a height of
  10~R$_*$.}
\label{f:sigmalx}
\end{figure}

The reason why $\dot\Sigma$ at radii in the 20-30 AU range appears to
be roughly the same for models with obscuring columns up to
$10^{20}cm^{-2}$ (which screen out EUV radiation) is that the thermal
structure of the disk at this radial distance is hardly affected by EUV 
radiation which is very weak at these radii due to large neutral H opacity of
the disk at EUV wavelengths. Radiation at soft X-ray 
wavelengths, on the other hand, is optimal for heating this region
of the disk.  

For columns larger than 10$^{20} cm^{-2}$ we see a drastic drop in total mass
loss rates, as shown in Table~1. A model with a pre-screening column of 10$^{21}cm^{-2}$
(FS21H2Lx1; cyan line in Figure~\ref{f:sigmah}) is
approximately a factor of 200 lower than for models with lower
pre-screening columns and a model with a $10^{22}cm^{-2}$ pre-screening column
(FS22H2Lx1; not plotted in Figure~\ref{f:sigmah})
shows no photoevaporation at all. This is due to the fact that these
high pre-screening column models, while having the same total X-ray luminosity
as less screened ones, have a much harder spectrum (see
Figure~\ref{f:jspectra}, but note that for clarity the spectra shown
in this figure have not been normalised). The reason for keeping the
total X-ray luminosity constant is to be able to isolate the effects
of the irradiating spectral shape. ``Hard'' X-rays ($\ga$
1~keV) have very  
large penetration depths and therefore they do not heat the upper
layers of the disk as efficiently as their lower energy
counterparts. Their energy is ``dispersed'' in a larger column and,
while these photons are vital to achieve a low-level ionisation deep
in the disk interior, which is a basic requirement of
the magneto rotational instability theory (Balbus and Hawley, 1991), their
contribution to the heating of the upper disk layers and therefore to
disk dispersal is insignificant. We note at this point that previous
work by Gorti \& Hollenbach (2009, GH09), who used a rather hard X-ray
spectrum to irradiate their disks also obtained very low
photoevaporation rates (see discussion in Section~\ref{s:comp}).

\paragraph{Height of the illuminating source} 
Results of models with the illuminating
source at a height H~=~2~$R_*$ are not too dissimilar from those with
H~=~10~$R_*$ (as assumed in Paper~I). The radial dependences of
$\dot\Sigma$ and \.M for a model with H~=~10~$R_*$ (FS0H10Lx1) are represented by the
dashed lines in the left and right panels of Figure~\ref{f:sigmah},
respectively. 
$\dot\Sigma$ at 5-10 AU for the H~=~10~$R_*$ model (FS0H10Lx1) is
comparable to the values 
obtained for the H~=~2~$R_*$ model (FS0H2Lx1), at $\sim$20 AU, however the H~=~10~$R_*$ model
produces about a factor of two higher $\dot\Sigma$, but negligible 
levels of photoevaporation at larger radii (30-40
AU). This results in  $\dot\Sigma \times 2 \pi R^2$ peaking at
$\sim$20 AU rather than $\sim$35~AU as for the model with 
H~=~2~$R_*$. The reason
for this inward shift of the peak mass loss flow is due to
the fact that radiation emitted from H~=~10~$R_*$ hits the inner disk
traveling along rays that are closer to the normal to the disk
surface, resulting in a more efficient heating of these regions. A
warmer inner disk is more puffed up and more likely to screen the material
at larger radii, hence producing lower photoevaporation rates than in
the H~=~2~$R_*$ case.  

The above interpretation is borne out by the fixed density structure
model results for H~=~10~$R_*$ and H~=~2~$R_*$ (FS0H10Lx1\_FG and
FS0H2Lx1\_FG). Here the density distribution of the  
disk is fixed and equal for both models and, indeed, we see no
difference in the mass loss rates (see Table~\ref{t:mods} and
dashed lines in Figure~\ref{f:sigmanh}).

We finally note that the total mass loss rate obtained from the
HSE model with H~=~10~$R_*$ (FS0H10Lx1)
is only about 20\% smaller than that obtained for the same model with
with H~=~2~$R_*$ FS0H10Lx1 (FS0H2Lx1), and
therefore not a real cause for concern, given the (larger) error
implicit to our $\rho c_s$ estimate of $\dot\Sigma$ (see Section~6).

\paragraph{X-ray luminosity}
Since X-ray luminosities of T~Tauri stars range over two orders of
magnitude or more, it is of interest to investigate the general trends
of photoevaporative mass loss rates with varying model X-ray
luminosity.  For a discussion of the connection between
photoevaporation, accretion rate and X-ray luminosities we refer to a
companion paper by Drake et al. (2009).

To eliminate the effects of EUV radiation, which may complicate any
trends due to X-rays, we irradiate the disk with the single
temperature X-ray model already used in Paper~I
(Figure~\ref{f:sp1}). Figure~\ref{f:sigmalx} shows the results for the
radial distribution of $\dot\Sigma$ and \.M for models with X-ray
luminosities varying between $2 \cdot 10^{29}erg/sec$ and $4 \cdot
10^{31}erg/sec$. 
The total mass loss rates are plotted as a function of X-ray
luminosity in Figure~\ref{f:lx}, which shows a roughly linear correlation
of these quantities. The increase in
X-ray luminosity means that  
deeper layers (higher density) in the disk can be heated to
temperatures in excess of the local escape temperatures, directly
resulting in an increase in the mass loss rates. In addition the
higher temperatures cause a vertical expansion of the heated surface
material, hence increasing the solid angle directly irradiated by the
source. However, the puffed up inner disk also obscures the disk at higher radii,
hence inhibiting further increases in mass loss rate. This explains the
slightly weaker than linear L$_X$-\.M relation in Figure~\ref{f:lx}
for high L$_X$. 
\begin{figure}
\plotone{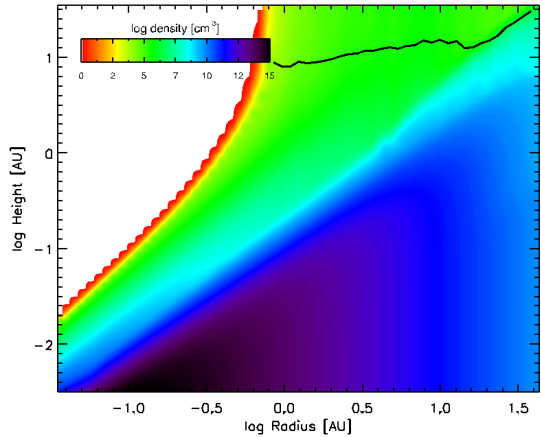}
\caption[]{Gas density distribution for the unscreened X-ray+EUV model
in HSE (FS0H0Lx1). The thick black line indicates the location of the
photoevaporative flow base.}
\label{f:2dhse}
\end{figure}

\begin{figure}
\plotone{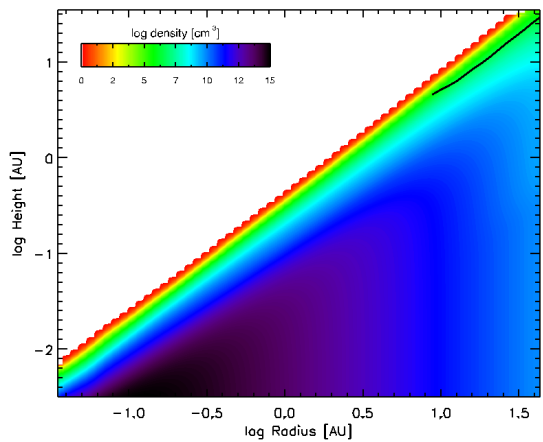}
\caption[]{Gas density distribution for the unscreened X-ray+EUV model
with fixed grids (FS0H0Lx1\_FG). The thick black line indicates the location of the
photoevaporative flow base.}
\label{f:2dnohse}
\end{figure}


\paragraph{Disk structure and Hydrostatic Equilibrium}
The results of models that use the HSE procedure
described in Section~\ref{s:hydro} are compared here with those
obtained using a fixed density structure for the same parameters. 

A casual inspection of Table~1 shows that hydrostatic
equilibrium models produce smaller mass loss rates than the fixed
density structure models with the same parameters. As in the case of
the height of the 
illuminating source, the difference is due to the fact that, unlike
fixed density structure models, heated material in the disk of HSE models puffs up and
screens material at larger radii from the source, producing lower
photoevaporation rates.

\begin{figure}
\plotone{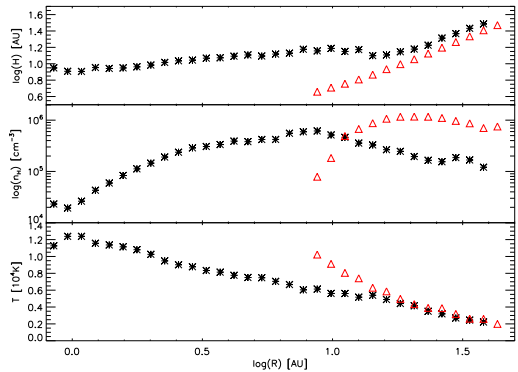}
\caption[]{Height, gas density and temperature of the photoevaporative
  flow surface for the unscreened X-ray+EUV models in HSE (black
  asterisks, FS0H2Lx1) and with fixed density (red triangles,
  FS0H2Lx1\_FG). At small cylindrical radii (R$\la$10\,AU) the
  temperatures required for the gas to escape are only reached at large
  heights above the midplane in the hydrostatic equilibrium
  models. The fixed grid models show no photoevaporation at all at
  R$\la$10\,AU due to the fact that the disk atmosphere is not allowed
  to expand and the escape temperatures are therefore never reached by
  the gas in these regions (see Section 3).}
\label{f:base}
\end{figure}

\begin{figure}
\plotone{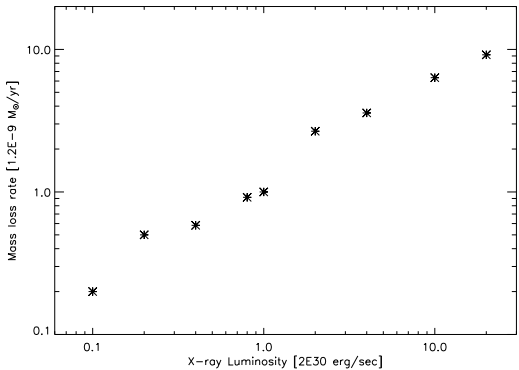}
\caption[]{Dependence of the total mass loss rate on X-ray luminosity.}
\label{f:lx}
\end{figure}

Figures~\ref{f:2dhse} and \ref{f:2dnohse} show the two-dimensional
density distribution of the disk for unscreened X-ray+EUV models
respectively in HSE and with a fixed density
(FS0H2Lx1 and FS0H2Lx1\_FG). The base of the photoevaporative envelope
is marked by the thick black line. The physical height, density and
temperatures of the flow surfaces are shown in Figure~\ref{f:base}. 

Figure~\ref{f:sigmanh} shows the radial distribution of $\dot\Sigma$
and \.M for the fixed density structure models. A comparison with
Figure~\ref{f:sigmah}, which shows the same quantities for the
HSE models, promptly reveals that, unlike fixed
density structure models, HSE models show 
  photoevaporation from radii smaller than $\sim$10~AU. This is 
due to the fact that heated gas in HSE models
is allowed to expand in the z-direction, moving away from the star to 
locations of lower gravitational potential where escape 
temperatures are lower, thus establishing a photoevaporative flow. 

Furthermore gas at larger radii in fixed density structure
models does not suffer
from attenuation from puffed up inner material, resulting in substantial
mass loss rates being supported in a larger portion of the disk. For
ease of computation, our simulations only included the inner 50AU of
the disk. As it is apparent from the left panel of
Figure~\ref{f:sigmanh}, this is not sufficient to estimate total mass loss
rates from low pre-screening ($< 10^{21} cm^{-2}$) fixed density structure
models, and the values of \.M given in Table~1 should be regarded as lower limits.

\section{Density and temperature structure} \label{s:ion}

A detailed discussion of the thermochemical and density structures
predicted by our models will be presented in future work, where predictions for gas emission line
diagnostics will also be compiled for a larger disk model. Nonetheless, it is helpful to
illustrate briefly here the effects of X-ray and EUV irradiation on the
density and 
temperature distribution of gas in the disk, in particular for those
regions where the photoevaporative flow is predicted to be significant. 

\begin{figure}
\plotone{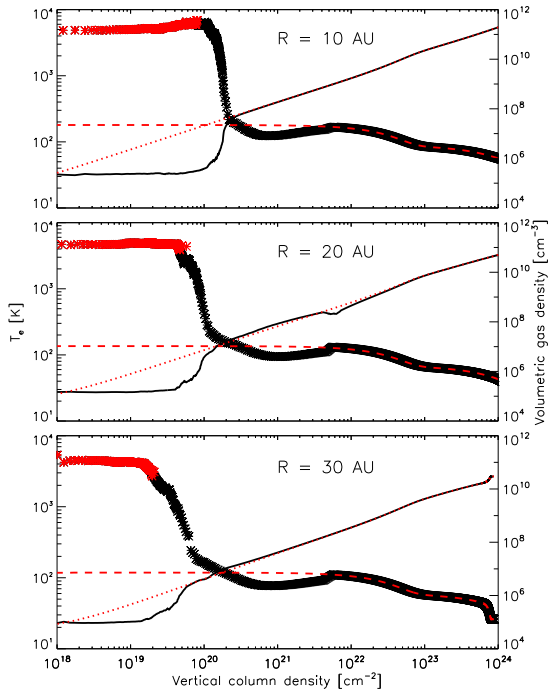}
\caption[]{Gas temperature (black asterisks) and density structure
  (solid line) for the unscreened
  X-ray+EUV model irradiated by a source at 2~R$_*$ (FS0H2Lx1). The
  red dashed line represents the dust temperature distribution of the
d'Alessio (2001) model. The red asterisks mark the locations where
the gas temperature exceeds the local escape temperature.}
\label{f:tempsfs0}
\end{figure}

Figure~\ref{f:tempsfs0} shows the density (black solid line) and
temperature structure (black asterisks) of the gas as a function of
vertical column density at radii $R = 10, 20$ and $30~AU$ for the
X-ray+EUV model with no pre-screening and $L_X = 2 \cdot 10^{30}erg/sec$
(FS0H2Lx1). The red dashed and the red dotted lines show the dust
temperatures and densities, respectively, calculated by d'Alessio et al. (1998).

The red asterisks indicate temperatures that are above the local
escape temperature, and therefore the base of the flow is located by the
red point at the highest column density. Since a small error on the
location of the flow could potentially translate into a large error on
the density (and therefore photoevaporation rates) great care was
taken to ensure that the Monte Carlo error is small in the region where the base
of the flow is located.  

The gas temperatures show a small step at columns of $\sim
5\cdot10^{21}cm^{-2}$. This is not realistic and an artifact of our
models which do not include a self-consistent treatment of molecular
chemistry. We impose the edge of the ``molecular zone'' from the
models of Glassgold et al. (2004), and assume that in the molecular
zone fine structure lines of neutrals are no longer efficient at
cooling the gas (since all gas is molecular). This causes the
temperature to rise slightly as the gas and dust become thermally
coupled. We stress that our temperature calculations in the deeper layers are
not reliable and we postpone further discussion to future work when a
chemical network will be introduced in the models. 


In the X-ray/EUV heated region the gas temperatures increase steeply
and the gas densities respond with an initial steep decline, however
both gas temperatures and densities flatten out at low column
densities. This temperature behaviour is expected for a photoionised
gas whose thermal balance in this region is dominated by heating by
photoionisation and cooling by collisionally excited line
emission. The density behaviour is due to the fact that at high z's,
where gravity is weak, the
exponential term in Equation~1 tends to unity and therefore the
density at given height is then only proportional to the ratio of the
gas temperature at the lower boundary point to the local gas
temperature, i.e. the hydrostatic structure is roughly isobaric.

This flattening of the densities at large heights demonstrates
somewhat a failure of the HSE assumption. In
reality, material in the upper disk atmosphere will be part of a flow
and not bound to the disk itself. This should not have a significant
effect on the photoevaporation rates, given that the isobaric layers
are not in the line of sight between the source and the flow surface.

\section{Comparison with previous work}
\label{s:comp}

Previous studies of the thermochemical properties and
photoevaporation of X-ray irradiated gas in protoplanetary disks
include ACP04, Glassgold et al. (2004, 2007), Ercolano et
al. (2008, Paper~I) and Gorti \& Hollenbach (2009, GH09). 

ACP04 used {\sc cloudy} (Ferland, 1996) (a one-dimensional
photoionisation code) to construct a simple 1+1D  
model to investigate the importance of X-ray photoevaporation for the
dispersal of gaseous protoplanetary disks around solar mass
stars. Using the simple 
$\rho c_s$ approximation they obtained a surface-normalized mass
loss rate
$\dot\Sigma =
2.6\cdot10^{-13} g\,s^{-1}cm^{-2}$ at a radial distance of
19.1~AU. Although their model was rather idealised and they used
slightly different input parameters for the central star and
illuminating spectral shape and luminosity, the values of 
$\dot\Sigma$ they
estimated is in the range of those obtained in this work (see
Figure~\ref{f:sigmah}), and corresponds to a total mass loss rate of
roughly 10$^{-9}~M_{\odot}/yr$. ACP04 compared their estimate of
$\dot\Sigma$ with that previously obtained for UV photoevaporation by
Hollenbach et al. (1994), and found comparable values
($\dot\Sigma_{UV} = 3.88\cdot10^{-13}g\,s^{-1}cm^{-2}$ at 6.7~AU),
which lead them to the conclusion that X-ray photoevaporation does not
play an 
important role. 
However, at a much larger radial distance of 19~AU---approximately
the region that 
Ercolano et al.\ (2008) found to dominate X-ray photoevaporation---ACP04 
obtained similar values of $\dot\Sigma$.
Due to the R$^2$
dependence of the total mass loss rates, this translates to a 
photoevaporation rate nearly an order
of magnitude higher than that at 6.7~AU. Therefore, while we agree
with the numerical results of ACP04, we interpret the result
differently and stress that our calculations (and those of ACP04) 
both suggest that X-ray photoevaporation does play a significant role in
disk dispersal. 

Glassgold et al. (2007) calculated the thermochemical structure
of a disk around a 0.5~$M_{\odot}$ pre-main sequence star, irradiated
by X-rays from a central source. The authors focused on the characterisation of
the physical properties of the disk and the prediction of gas-phase
diagnostics and do not give an estimate for X-ray photoevaporation
from their models. However from the model grids which were kindly
made available to us by these authors, we find $\dot\Sigma =
2.2\cdot10^{-12} g\,s^{-1}cm^{-2}$ at 20~AU. While this is higher than
the estimates of this work and ACP04, it agrees very well with the
values reported in Paper~I. 
We note that Glassgold et al. (2007) and Paper~I
both used a fixed density structure
for their calculations, and the higher
photoevaporation rates are due to the fact that no obstruction by the
puffed-up inner disk is accounted for in these models. 

Contrary to this work and to the results mentioned above, recent work
by Gorti \& Hollenbach (2009, GH09) predicts much smaller
photoevaporation rates due to direct X-ray heating and concludes that X-rays are only of indirect importance for
the dispersal of protoplanetary disks around solar mass stars. 
Instead, their models predict that X-ray ionization generally acts to ``amplify'' FUV photoevaporation rates. While the authors state that their results are in agreement with those
of ACP04, inspection of their Figure~5 (curve labeled $X$) shows that
for their ``X-ray only'' model they estimate $\dot\Sigma$ of the order of
10$^{-15} g\,s^{-1}cm^{-2}$ at 20~AU, which is
two orders of magnitude smaller than
our and ACP04's estimates. 
We also note that ACP04 obtained their estimate of $\dot\Sigma \sim
10^{-13} g\,s^{-1}cm^{-2}$ at a radius of 19.1~AU, rather than 10~AU as
stated in the GH09 paper.  

Comparison of complicated theoretical calculations can be
difficult and the reason for disagreement may be buried in the codes 
themselves or in the physical assumptions adopted. 
However we have identified at least one reason for the discrepancy between our calculations and those of GH09 which is the hardness of the illuminating X-ray spectrum. 

Indeed, we find that the soft-X-ray region is the most efficient
at heating the gas in the photoevaporation layer. GH09 use a rather
hard spectrum, which peaks at 2keV and is approximated by a power-law
with dL$_X$/dE $\sim$ E for $0.1 < E < 2$ keV and dL$_x$/dE $\sim$ E$^{-2}$ for 
$2 < E < 10$ keV. This is a significantly harder spectrum than that
used by ACP04 (which peaked at 0.7keV), ourselves in Paper~I (see
Figure~\ref{f:sp1}) and by Glassgold et al. (2007, which peaked at
1keV). In this work we found that models irradiated by a hard spectrum
such as that produced by high column screens (e.g. FS22 models) did
not yield significant photoevaporation.  

We have run a model using the GH09 spectrum and we obtained
$\dot\Sigma \sim 6\cdot10^{-14} g\,s^{-1}cm^{-2}$ at 20~AU (using the
$\rho c_s$ approximation) and a total mass loss rate of
$\sim$10$^{-10}~M_{\odot}/yr$. While a step in the 
right direction, the hardness of the illuminating spectrum alone does
not seem to be sufficient to explain the discrepancy and other factors must also
come into play. 

\section{Limitations of current models}
\label{s:lims}

\subsection{The estimation of photoevaporation rates from the models}

 By far the biggest uncertainties concern our attempt to use these hydrostatic
radiative equilibrium models to estimate the resulting mass loss rates.
This now becomes a critical problem given that our preliminary estimates
suggest that, contrary to previous claims (ACP04, GH09), X-ray photoevaporation
{\it may} be the major agent of disc dispersal.

The uncertainties stem  from the fact that we have derived the modified
disc structure under the assumption of HSE, and although
this probably gives a reasonable description of the density structure at small
z, it is obviously the case that a {\it static} description does not
produce a corresponding mass {\it flow} rate. Instead, we have employed the
simple approach of deriving the hydrostatic structure and then, at the height
where the disc material exceeds the local escape temperature, assume free
sonic outflow from that point. A similar approach in the EUV case is less
questionable for two reasons: i) the steep temperature discontinuity at
the ionisation front and the near isothermal conditions above it
means that the location of the launch surface is independent of detailed
modeling and just depends on the regions of the disc where
photoionised gas is unbound and ii) the density at the launch surface
depends only on conditions above the ionisation front and not on the
structure of the underlying disc. Thus the fact that 
the disc below the launch surface is not strictly
static has no effect on the resulting outflow rates.
 
  In the X-ray case, the temperature rises much more gradually with height:
for example, for our model at 20 AU, the temperature rises from roughly
$200$ to $5000$ K over a region about 6.5 AU in height, over which the density of the
hydrostatic structure falls by one order of magnitude. In reality, pressure
gradients will cause the
flow to  be initiated below our nominal launch point, i.e. at a height
where its temperature is less than the
escape speed, and the density structure will be progressively modified
over the region where the flow is accelerated. It therefore becomes little
more than an exercise in educated guesswork to assign a point for estimating
the escaping mass flux - the lower down in the structure one tries to do this,
the more reliably the density structure will be known, but the less easy it is
to assign a flow speed. At greater heights, it's easier to assign a flow speed
(unbound gas expands at about the sound speed), but the estimate of
local density becomes unreliable, as the underlying density structure is
now considerably modified by the flow.

   It is tempting (see e.g. GH09) to propose a simple, physically motivated
prescription that maps the structure of the photo-heated region onto an
analytic flow solution at large radius (for example, the isothermal Parker
wind solution). The problem with this is that although at large
radii the flow will indeed approximate a quasi-spherical radial outflow,
the transition between the nearly hydrostatic structure at the base and the
Parker-like structure at large radius is quite different from a Parker wind -
in particular, when the flow is non-radial, the interplay between terms
in the momentum equation involving the divergence of the velocity, the
gravitational acceleration and the centrifugal acceleration is quite
different from the non-rotating Parker case. The problem boils down to
the fact that, until one knows the topology of the streamlines, one cannot
compute the variation of quantities along the streamlines -  but that the 
steady state topology of the streamlines itself depends on the variation
of quantities along the streamlines: in other words, it is a truly
two dimensional problem. (See Begelman et al., 1983 for an attempt to
parameterise the problem in terms of an ad hoc variation of the divergence
of the velocity along streamlines). Our initial hydrodynamic investigations
of this problem (Alexander \& Clarke in prep.) demonstrate that the 
streamline topology and the mapping between base density and mass flux
is very dependent on how the base density varies with radius in the disc. 
We therefore think that it is presently unwarranted to try and go beyond
the simple estimates employed here and that further progress in estimating
photoevaporation rates will require a combined hydrodynamics/radiative
transfer approach. 

A further shortcoming is that  
a molecular chemistry network is still not included in our
calculations. This poses a limitation on the accuracy
of our gas temperature calculations in the lower layers
where cooling 
by H$_2$ and CO rovibrational and rotational lines may become
important. For example at a radial distance of 1~AU and column densities of
10$^{21}$-10$^{22}$~cm$^{-2}$ Glassgold et al. (2004) find that CO
rovibrational lines  dominate disk cooling.  CO rotational lines
were also found to contribute somewhat 
to the cooling at greater depths, although
in these regions the thermal balance is dominated by dust-gas
collisions (GNI04). We are currently in the process of developing a
chemical network to be included in future studies and will therefore
be able to derive a more accurate thermal/hydrostatic structure at low
temperatures ($\sim 10^2$). It is questionable, however, how much such
improvements will affect our estimates of photoevaporation rates, given
that gas in this temperature range is strongly bound at radii less
than 100 AU.

\section{Summary}\label{s:summary}

  Our calculations of the structure of X-ray irradiated discs in low
mass pre-main sequence stars imply total photoevaporation rates that, taken
at face value, exceed the photoevaporation rates produced by extreme
ultraviolet (EUV) radiation by an order of magnitude. This difference
is attributable to the fact that a much more extended region of the disc
is subject to X-ray photoevaporation than in the EUV case (i.e. significant
evaporation out to $\sim 40$ AU, in contrast to the EUV case where little
mass is lost beyond $\sim 10$ AU). The reason for this different
photoevaporation profile is that the X-rays are able to penetrate
columns of up to $\sim 10^{22}$cm$^{-2}$ through the disc's puffed up,
X-ray irradiated atmosphere 
and thus still produce
significant mass flows when they impact the disc at several tens of AU.
In the EUV case, by contrast, the regions of the disc from which
the flow can be launched are shielded from {\it direct}
irradiation by stellar photons and are instead subject only to
recombination photons from
the bound region of the disc's ionised atmosphere. Since gas at $10^4$K is
only bound out to $\sim 5$ AU, most of the EUV induced
mass loss is restricted  
to regions modestly greater than this (Hollenbach et al., 1994,
Richling \& Yorke, 2000; Font et al, 2004). Our conclusion that X-ray photoevaporation
is more important than EUV photoevaporation is only strengthened when
one considers two additional factors: i) the possibility of
further attenuation of the EUV flux by absorption by any neutral flows
(such as winds) close to the star and ii) the fact that we have unambiguous
evidence that young stars remain strong X-ray sources even after disc dispersal
and thus are clearly available as a potential disc dispersal mechanism.

  However, we emphasise that our estimates of X-ray photoevaporation rates
are {\it much} more uncertain than the EUV values, notwithstanding our
careful treatment of two-dimensional radiative transfer and our 
derivation of the puffed up density structure of the irradiated disc
(see discussion in Section 6). These uncertainties were of academic
interest as long as it seemed likely that X-ray photoevaporation was in any
case a minor agent for disc dispersal. We argue that future radiation
hydrodynamical simulations are required in order to make firm predictions
about the role of X-ray photoevaporation in disc dispersal.

\section*{Acknowledgments}
We thank Uma Gorti and David Hollenbach for useful discussion. We
thank Al Glassgold for providing the data grids for their model. We 
thank John Raymond for helpful discussions and a critical assessment of this work.
We thank Paola D'Alessio for providing us with the electronic data for
the gas density distribution in the disk. JJD was supported by the
Chandra X-ray Center NASA contract NAS8-39073 during the course of
this research.  This work was performed using the Darwin Supercomputer
of the University of Cambridge High Performance Computing Service
(http://www.hpc.cam.ac.uk/), provided by Dell Inc. using Strategic
Research Infrastructure Funding from the Higher Education Funding
Council for England.

\end{document}